# Extraordinary *Spin specific beam shift* of Light in an Inhomogeneous Anisotropic medium

Mandira Pal, Chitram Banerjee, Shubham Chandel, Ankan Bag, and Nirmalya Ghosh[*]

*Dept. of Physical Sciences, IISER- Kolkata, Mohanpur 741 252, Nadia, West Bengal, India*

*E-mail: **nghosh@iiserkol.ac.in**

Nirmalya Ghosh

Department of Physical Sciences,

Indian Institute of Science Education and Research (IISER) – Kolkata, Mohanpur Campus,

Mohanpur, West Bengal, India, 741252

Phone:  +91- 03473 279130, Ext- 206 (Office); +91-9734 678247 (Cell)

Fax: +91 33 2587 3020

**ABSTRACT**

Spin orbit interaction and the resulting Spin Hall effect of light are under recent intensive investigations because of their fundamental nature and potential applications. Here, we report an extraordinary *spin specific* beam shift of light and demonstrate its tunability in an inhomogeneous anisotropic medium exhibiting spatially varying retardance level. The spin specificity (shift occurs only for one circular polarization mode, keeping the other orthogonal mode unaffected) is shown to arise due to the combined spatial gradients of the geometric phase and the dynamical phase of light. The constituent two orthogonal circular polarization modes of an input linearly polarized light evolve in different trajectories, eventually manifesting as a large and tunable spin separation. The spin specificity of the beam shift and the demonstrated principle of simultaneously tailoring space-varying geometric and dynamical phase of light for achieving its tunability (of both magnitude and direction), may provide an attractive route towards development of spin-optical devices.

Spin orbit interaction (SOI) dealing with the coupling of spin and orbital degrees of freedom of massive and mass-less particles has led to several fundamental consequences in diverse fields of physics, ranging from atomic, condensed matter to optical systems. SOI of light refers to the coupling between the spin (SAM, circular / elliptical polarization) and orbital angular momentum (OAM, phase vortex) degrees of freedom [1-5]. This is associated with two interdependent effects – *(i)* evolution of geometric phase due to the effect of the trajectory on the state of polarization of light, leading to intrinsic SAM to intrinsic OAM inter-conversion and its various intriguing manifestations (such as formation of polarization controlled vortices); and *(ii)* the reverse effect of polarization on the trajectory of light, leading to intrinsic SAM to extrinsic OAM inter-conversion and manifesting as the so-called Spin Hall effect (SHE) of light [1-5]. The photonic SHE has been observed recently in various optical interactions [1-5], each of which are discernable by important fundamental or applicative aspects. These are under recent intensive investigations because of their fundamental nature and also due to the fact that these are offering new opportunities for the development of spin-controlled photonic devices [1,2, 5, 6]. The rather weak SOI effect and the exceedingly small magnitude of SHE (typically in the sub-wavelength domain) is a major stumbling block towards their practical applications [1-3]. Despite considerable recent efforts towards enhancing these effects in diverse optical systems [3, 5, 7, 8], realization of tunable spin-dependent splitting of light beam remains to be an outstanding challenge. Here, we report a new variant of photonic SHE, namely, the *spin specific* beam shift and demonstrate its full tunability in an inhomogeneous anisotropic medium exhibiting user-controlled spatially varying retardance level. The extraordinary spin specificity (shift occurs only for one circular polarization mode, keeping the other orthogonal mode unaffected) is shown to arise due to the combined spatial gradients of the geometric and dynamical phases of light. In a simple yet elegant system of a twisted nematic liquid crystal-based spatial light modulator (SLM), we demonstrate that one can simultaneously generate desirable spatial gradients of both the geometric and the dynamical phases of light to produce spin specific beam shift in a regulated fashion. The effect is eventually manifested as a spin dependent splitting of input linearly polarized beam, where the constituent two orthogonal circular polarization modes evolve in different trajectories resulting in a large and tunable spin separation. The *spin specificity* of the beam shift and the demonstrated principle of simultaneously tailoring space-varying geometric and dynamical phase of light for achieving its tunability (of both magnitude and direction), may provide an attractive route towards development of spin-optical devices for spin-controlled photonic applications [6].

*Spin Specific Beam shift*

Temporal evolution of geometric phase is known to manifest as an input polarization-dependent shift of the frequency (ω) of light [9-11]. Such an effect takes place when circularly polarized light passes through a rotating homogeneous anisotropic medium (e.g, half waveplate), the resulting Pancharatnam-Berry (PB) geometric phase evolves linearly in time ($\Phi_g(t) \approx 2\Omega t, \Omega$ is the rotation rate), manifesting as an input circular polarization (SAM)-dependent frequency shift $\Delta\omega = \pm \frac{d\Phi_g}{dt} = \pm 2\Omega$ (± corresponding to input right (RCP) / left (LCP) circular polarization states) [10]. The spatial analogue of this effect in a transversely inhomogeneous anisotropic medium is related to the SOI of light [12-14]. Here, we describe a simple yet intriguing effect associated with the temporal↔spatial analogy of the geometric phase gradient. A Gaussian beam propagating along the z-direction of an inhomogeneous (along the transverse x/y direction) anisotropic medium is associated with transverse momentum components *($k_\perp=k_x$ and $k_y$)*. A space varying PB geometric phase $\left(\Phi_g(\xi) \approx \Omega_\xi \xi; d\Phi_g/d\xi=\Omega_\xi; \xi \rightarrow x/y\right)$ should manifest as SAM -dependent shift in the transverse momentum distribution *($\Delta k_\perp$)* of the beam (temporal frequency↔spatial frequency). If one further introduces a spatial gradient of dynamical phase $\left(\Phi_d(\xi) \approx \Omega_\xi \xi; \frac{d\Phi_d(\xi)}{d\xi} = \Omega_\xi\right)$, it may happen that for one circular polarization mode, the two spatial gradients cancel out to yield no net shift. For the other orthogonal mode, on the other hand, they may yield an accumulated shift of the beam centroid. We refer to this effect as the spin specific beam shift, as formulated below.

For a paraxial Gaussian beam, we neglect small longitudinal field component and represent the transverse components as

$$|\mathbf{E}_i\rangle = |\mathbf{e}\rangle F(x,y) \qquad (1)$$

where $|\mathbf{e}\rangle = [\alpha_1, \alpha_2]^T$ is the Jones vector of the homogeneously polarized input beam and $F(x,y)$ is its Gaussian envelop. On propagation through the inhomogeneous anisotropic medium, it acquires space varying dynamical ($\Phi_d(\xi)$) and PB geometric phase ($\Phi_g(\xi)$). The output field can be written as

$$|\mathbf{E}_o\rangle = e^{i\phi_d(\xi)} J_{PB}(\xi)|\mathbf{E}_i\rangle \qquad (2)$$

Here, $J_{PB}(\xi)$ is a 2×2 matrix representing the effect of the PB geometric phase $\Phi_g(\xi)$ [14].

For input RCP ($|\mathbf{e_R}\rangle = \frac{1}{\sqrt{2}}[1, i]^T$) and LCP ($|\mathbf{e_L}\rangle = \frac{1}{\sqrt{2}}[1, -i]^T$) states, the output field becomes

$$|\mathbf{E_o}\rangle = e^{i\{\phi_d(\xi) \pm \phi_g(\xi)\}}|\mathbf{E_i}\rangle \qquad (3)$$

Here, ± correspond to input RCP and LCP states, respectively.

The expectation values of the transverse co-ordinates and the momenta of the output beam can then be calculated as

$$\langle \xi = x/y \rangle = \frac{\langle E_0|\xi=x/y|E_0\rangle}{\langle E_0|E_0\rangle} \;; \quad \langle k_{\xi=x/y}\rangle = \frac{\langle E_0|i\frac{\partial}{\partial \xi=x/y}|E_0\rangle}{\langle E_0|E_0\rangle} \qquad (4)$$

For input RCP and LCP states, $\langle \xi = x/y \rangle$ and $\langle k_{\xi=x/y}\rangle$ can be determined by using equation (3) in equation (4). While, $\langle \xi = x/y \rangle$ vanishes, yielding no net co-ordinate shift, the momentum shift becomes non-zero and determined by $\Phi_d(\xi)$ and $\Phi_g(\xi)$. When the two phases have equal gradient ($\frac{d\Phi_d(\xi)}{d\xi} = \frac{d\Phi_g(\xi)}{d\xi} = \Omega_\xi$), the momentum domain shifts for input RCP and LCP states become

$$\langle k_{\xi=x/y}\rangle_{RCP} = 2\Omega_\xi \quad and \quad \langle k_{\xi=x/y}\rangle_{LCP} = 0 \qquad (5)$$

We shall subsequently demonstrate that equal spatial gradients of the PB geometric phase and dynamical phase of light can indeed be generated in an inhomogeneous anisotropic medium to produce the spin specific beam shift. Such an inhomogeneous anisotropic medium can be realized by modulating the pixels of a twisted nematic liquid crystal-based SLM by user controlled grey level distributions. We now turn to modeling evolution of PB geometric phase and dynamical phase of light in such system.

*Pancharatnam-Berry (PB) geometric phase and dynamical phase in twisted nematic liquid crystal layers*

The evolution of both the phase and the polarization of light in twisted nematic liquid crystal layers can be modeled using the following equation [15]

$$|\mathbf{e_o}\rangle = e^{i\left(\frac{2\pi n_o d}{\lambda} + \frac{\delta_{tot}}{2}\right)} R(-\psi) \begin{pmatrix} A - iB & C \\ -C & A + iB \end{pmatrix} |\mathbf{e_i}\rangle = e^{i\left(\frac{2\pi n_o d}{\lambda} + \frac{\delta_{tot}}{2}\right)} J(\psi, \delta_{tot})|\mathbf{e_i}\rangle$$

Where, $A = \cos\left(\sqrt{\psi^2 + \left(\frac{\delta_{tot}}{2}\right)^2}\right)$; $B = \frac{\delta_{tot}}{2} \frac{\sin\sqrt{\psi^2 + \left(\frac{\delta_{tot}}{2}\right)^2}}{\sqrt{\psi^2 + \left(\frac{\delta_{tot}}{2}\right)^2}}$; $C = \psi \frac{\sin\sqrt{\psi^2 + \left(\frac{\delta_{tot}}{2}\right)^2}}{\sqrt{\psi^2 + \left(\frac{\delta_{tot}}{2}\right)^2}}$ (6)

Here, $\delta_{tot} = \frac{2\pi}{\lambda}(n_e - n_o)d$ is the total linear retardance, $n_e$ and $n_o$ are the extraordinary and ordinary refractive indices, $\psi$ is the twist angle, $J$ is the Jones matrix of the system containing the 2×2 rotation matrix $R(-\psi)$. The evolution of polarization can also be alternatively modeled using the effective Jones matrix ($J_{eff}$) as a sequential product of matrices of an equivalent linear retarder ($J_{reta}$, with effective linear retardance $\delta_{eff}$ and its orientation angle $\theta_{eff}$) and an effective optical rotator (with optical rotation $\psi_{eff}$) [16]

$$J_{eff} = R(\psi_{eff}) J_{reta}(\delta_{eff}, \theta_{eff})$$
$$\text{with} \quad \psi_{eff} = -\psi + 2\theta_{eff} \qquad (7)$$

The dynamical phase (for input LCP /RCP states) is clearly given by the phase factor in equation 6

$$\Phi_d(n) = \frac{2\pi}{\lambda}\left\{\frac{n_e(n)+n_o}{2}\right\}d = \frac{\delta_{tot}(n)}{2} + \frac{2\pi}{\lambda}n_o d \qquad (8)$$

Here, $n$ is the value of grey level(s) applied to a twisted nematic liquid crystal-based SLM. Henceforth, we only consider the first term in equation (8), since, the grey level dependent part is only relevant to the effect. The matrix $J$ (in equation 6) is free from dynamical phase $[\Phi_d = arg\left\{\frac{det(J)}{2}\right\} = 0]$ [17] and the corresponding PB geometric phase encoded in it, can be determined using the Pancharatnam connection [17-19], for input RCP ($|e_R\rangle$) and LCP ($|e_L\rangle$) states as

$$\Phi_g^+(n) = arg\langle e_R|J|e_R\rangle = +\psi_{eff}(n) \text{ and } \Phi_g^-(n) = arg\langle e_L|J|e_L\rangle = -\psi_{eff}(n) \qquad (9)$$

The origin of the geometric phase worth a brief mention here. This arises here due to non-cyclic polarization evolution [17, 19] in the twisted birefringent structure, wherein both the magnitude of birefringence and its varying orientation (due to twist along the longitudinal direction) contribute to its evolution.

Note that equation (7), (8) and (9) in combination with experimental Mueller matrix [20,21] measurements, can be used to determine both the geometric and the dynamical phases of light. For this purpose, full 4×4 Mueller matrices $M$ can be recorded from the SLM having uniform grey level ($n$) addressing. The effective linear retardance $\delta_{eff}(n)$ and optical rotation $\psi_{eff}(n)$ parameters can be determined from the elements of $M$, by representing it as a product of basis matrices of an equivalent linear retarder and rotator (Jones → Mueller matrix conversion of equation 7) (see Supplementary information) [21]. The twist angle of the SLM ($\psi$) can also be determined separately [16]. Using these set of parameters, one can determine the magnitudes of total retardance $\delta_{tot}(n)$ from the relationship connecting them (derived from the equivalence of equation 6 and 7) [16].

$$\delta_{tot}(n) = 2\sqrt{\left[\cos^{-1}\left(\cos\left(\frac{\delta_{eff}(n)}{2}\right)\cos(\psi_{eff}(n)+\psi)\right)\right]^2 - \psi^2} \qquad (10)$$

Thus obtained $\delta_{tot}(n)$ and $\psi_{eff}(n)$ parameters may finally be used to determine the values of $\Phi_d(n)$, $\Phi_g^{\pm}(n)$, and the resulting total phase [$\Phi_{tot}^{\pm}(n) = \Phi_d(n) + \Phi_g^{\pm}(n)$, ± corresponding to input RCP and LCP states, respectively].

In what follows, *(i)* we experimentally demonstrate extraordinary spin specificity of the beam shift and its tunability in an inhomogeneous anisotropic medium; *(ii)* We then determine both the space varying dynamical and the geometric phases in such system to demonstrate the underlying principle; *(iii)* Finally, we show large spin dependent splitting of input linearly polarized beam.

A schematic of the experimental system is shown in Figure 1. Fundamental Gaussian (TEM$_{00}$) mode of 632.8 nm line of a He–Ne laser (HRR120-1, Thorlabs, USA) is spatially filtered, collimated (using Lens, Pinhole and aperture assembly), and made incident on a transmissive SLM (LC-2002, Holoeye Photonics, 832 × 624 square pixels, pixel dimension 32 μm). The transverse momentum distribution ($k_\perp = k_x$ and $k_y$) of the transmitted beam is imaged into a CCD camera (Micro Publisher 3.3, Qimaging, 2048 × 1536 square pixels, pixel dimension 3.45μm). The polarization state generator (PSG) and the polarization state analyzer (PSA) units are used to generate and analyze desirable polarization states. The PSG unit comprises of a fixed Glan-Thompson linear polarizer (P$_1$, GTH10M, Thorlabs, USA) and a rotatable quarter waveplate (QWP$_1$, WPQ10M-633, Thorlabs, USA) mounted on a computer-controlled rotational mount (PRM1/MZ8, Thorlabs, USA). The PSA unit essentially consists of a similar arrangement of a linear polarizer (P$_2$) and a quarter waveplate (QWP$_2$), but positioned in a reverse order. In order to observe tunable spin specific beam shift and the spin-dependent splitting of light beam, desirable grey level (*n*) gradient in the SLM pixels was created along one chosen linear direction (x or y) using a range of grey level values between *n* = 30 to 170 (this choice was driven by the observed gradual variation of total phase experienced by the RCP state $[\Phi_{tot}^{+}(n) = \Phi_d(n) + \Phi_g^{+}(n)]$ with *n* for this range, shown subsequently in Figure 3c). Variable spatial gradient of grey levels was achieved by accommodating this within variable spatial dimensions (2.144mm – 6.816 mm) of the relayed image to the SLM. While studying the spin specific beam shift, the PSA unit was removed and the PSG unit was used to sequentially generate RCP and LCP polarization states. For observing the spin-dependent splitting of light, the PSG was used to generate linear polarization state, and the transmitted beam was sequentially analyzed for RCP and LCP analyzer basis states of the PSA.

The same system was also used to record the Mueller matrices of the SLM having different uniform grey level addressing. The Mueller matrix measurement strategy is based on sequential generation (using the PSG unit) and analysis (by the PSA unit) of four optimized elliptical polarization states [22, 23]. A series of sixteen intensity measurements (images) were performed by sequentially changing the orientation of the fast axis of the quarter waveplates of the PSG unit and that of the PSA unit, to four optimized angles 35°, 70°, 105° and 140° (see Supplementary Information). The axes of $P_1$ and $P_2$ were fixed along the laboratory horizontal and vertical directions respectively. These sixteen intensity measurements were then combined to generate the system Mueller matrix following the approach described in [23].

*Tunable Spin Specific Beam Shift in inhomogeneous anisotropic medium*

Making use of varying spatial gradient of the grey levels in the SLM, spin specific beam shift (shift in transverse momentum distribution $\frac{k_{x/y}}{2\pi}$) and its tunability is demonstrated in Figure 2. The momentum domain beam shift for the input RCP state increases systematically with increasing spatial gradient of grey level ($\frac{dn}{d\xi=x/y}$). Remarkably, the shift of the beam centroid (momentum shift manifested as a shift of the centroid) becomes as large as ~ 25.16 μm for the highest applied spatial gradient $\frac{dn}{d\xi}$= 0.0653 bit/μm (varied between 11.56 -25.16 μm for $\frac{dn}{d\xi}$= 0.0205-0.0653 bit/μm). The beam centroid for input LCP state, on the other hand, does not exhibit any appreciable shift (Figure 2). While, these results are for applied grey level gradient along the x-direction, similar results were also obtained for that applied along the y-direction. We now proceed to determine the space varying PB geometric phase and dynamical phase and relate these to the observed effect (the values of the spatial gradients of total phase $\frac{d\Phi_{tot}^{+}(x)}{dx}$ for the RCP state noted in Figure 2, are based on these, as described below).

*Experimental determination of space varying PB geometric phase and dynamical phase*

The geometric and the dynamical phases of light were determined from experimental Mueller matrix measurements in combination with the treatment outlined previously. Mueller matrices were recorded from the SLM having different uniform grey level (*n*) addressing (see Supplementary information). The twist angle of the SLM was separately determined to be $\psi = \pi/2^{16}$. The results of the Mueller matrix measurements, determination of the medium polarization parameters [$\delta_{eff}(n)$, $\psi_{eff}(n)$ and $\delta_{tot}(n)$], and the corresponding dynamical ($\Phi_d(n)$) and geometrical phases ($\Phi_g^{\pm}(n)$) of light are summarized in Figure 3. A typical Mueller matrix recorded from the SLM for *n* = 120 is shown in Figure 3a, where the elements are represented in normalized unit (normalized by the $M_{11}$ element). The normalized values of the elements (between -1 to +1) are shown using the color bar. The Mueller matrices show characteristic features of pure retarders (Figure 3a, and Supplementary information). Weak magnitudes and negligible variation of the elements of the 1st row and the 1st column of the matrices (which encode diattenuation effect [20, 21]) with varying *n* implies negligible polarization dependent amplitude modulation effect. The estimated medium polarization parameters ($\delta_{eff}(n)$, $\psi_{eff}(n)$ and $\delta_{tot}(n)$, Figure 3b) and the resulting total phase ($\Phi_{tot}^{\pm}(n) = \Phi_d(n) + \Phi_g^{\pm}(n)$, derived using equations 8, 9 and 10, Figure 3c) underscore the key feature pertinent to the observed spin specific beam shift. While for input RCP state, $\Phi_{tot}^{+}(n)$ increases gradually with increasing *n* (for the range $n \approx 30 - 170$ used in the experiments and accordingly displayed in Figure 2*)*, the corresponding variation for input LCP state $\left(\Phi_{tot}^{-}(n)\right)$ is rather weak and negligible. Comparison of the experimental momentum domain beam shifts for the input RCP state and the corresponding theoretical predictions (using the results of Figure 3c in equation 5) shows reasonable agreement (Figure 3d). For the theoretical predictions, an approximated linear dependence of $\Phi_{tot}^{+}(n)$ with *n* (for $n \approx 30 - 170$ in Figure 3c) was assumed and the spatial dimensions (over which the grey levels were applied) were duly considered (the values of the spatial gradients $\frac{d\Phi_{tot}^{+}(x)}{dx}$ noted in Figure 2 are based on this approximation). Incorporation of the exact dependence of $\Phi_{tot}^{+}(n)$ did not lead to significant differences in the predicted trends and absolute values. These results provide concrete evidence of the underlying principle– while for input RCP state, the accumulation of the spatial gradients of the dynamical and the geometrical phases lead to a large shift of the beam-centroid, the spatial gradients nearly cancel out to yield no appreciable shift for the input LCP state.

*Spin dependent splitting of input linearly polarized light beam*

The spin specificity of the beam shift is eventually manifested as a spin dependent splitting of input linearly polarized beam. Like in photonic SHE, the constituent two orthogonal circular polarization modes evolve in different trajectories, leading to a large (and tunable) spin separation (shown in Figure 4a and 4b). The effect is however, perfectly discernible from SHE in that – *(a)* only one circular polarization mode experiences the shift, the other orthogonal mode evolves in the same trajectory, *(b)* the magnitude as well as the direction of the splitting is completely tunable. Moreover, unlike other variant of the momentum domain beam shifts (angular Goos-Hänchen and the angular Imbert-Federov shifts [8, 24]), the shift is independent of the beam waist parameter and is exclusively determined by the dynamical and the geometric phase gradients.

To summarize, we have observed an extraordinary *spin specific* momentum domain beam shift of light and demonstrated its tunability in an inhomogeneous anisotropic medium. The effect is manifested as a shift of the beam centroid for one circular polarization mode whereas the other orthogonal mode remains unaffected and evolves in the same trajectory. This is shown to arise due to the combined spatial gradients of PB geometric phase and dynamical phase of light. It is pertinent to emphasize here that the SOI and the SHE effects in inhomogeneous anisotropic medium usually deal with systems having spatially varying axis of retardance (and with constant magnitude of retardance) [7,13,25,29]. In contrast, as demonstrated here, if one introduces a spatially varying magnitude of retardance, one may simultaneously generate space varying PB geometric phase and dynamical phase of light in a regulated manner to produce spin specific beam shift. While the effect is demonstrated in a twisted nematic liquid crystal-based system, albeit with a relatively smaller spatial gradient, the principle can be extended to a wide class of anisotropic nanooptical systems [5,6, 25-27], wherein the phase gradients can be enhanced by several orders of magnitude to produce giant spin specific beam shift. For example, space varying anisotropy effects can be tailored in specially designed plasmonic nano structures [26-30] to produce such effects. We are currently expanding our investigations in this direction. In general, the remarkable simplicity of the approach of simultaneously tailoring spatial gradient of geometric and dynamical phases of light to produce such dramatic spin specific beam shift may provide an attractive route towards development of spin-controlled photonic devices for the generation, manipulation and detection of spin-polarized photons.

**Figures:**

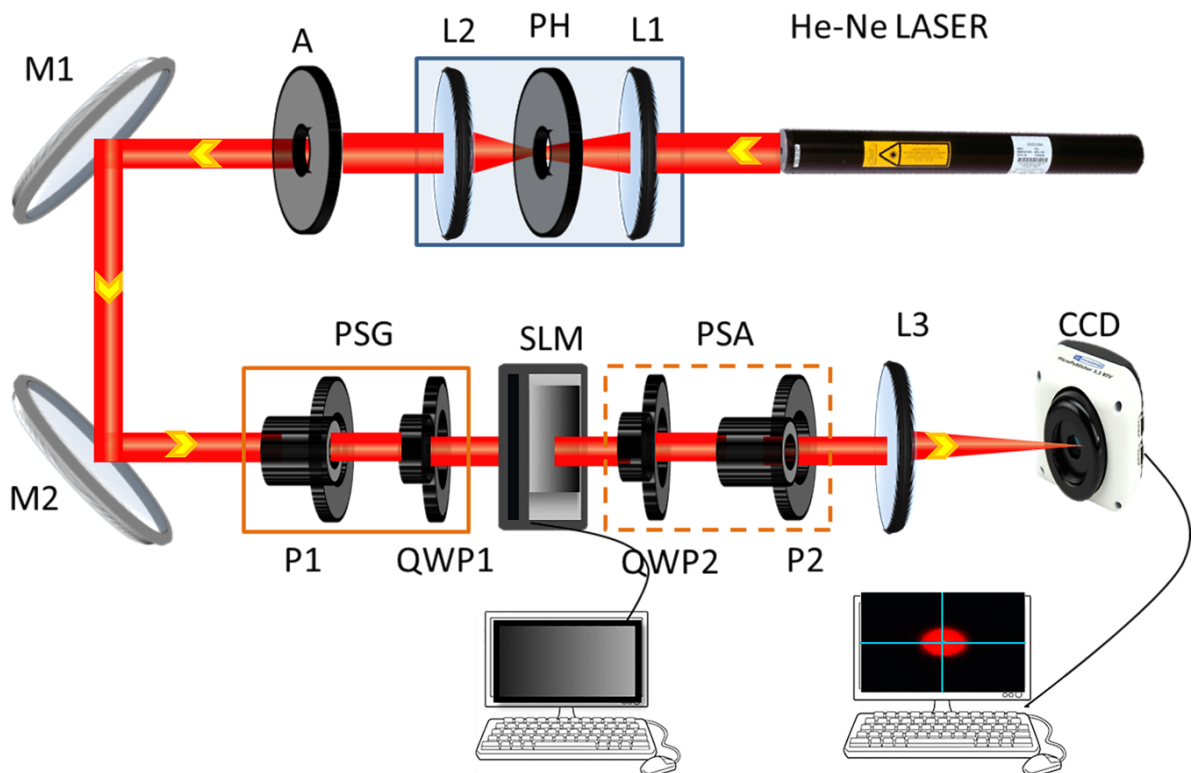

**Figure 1:** Schematic of the experimental arrangement for observing tunable spin specific beam shift and spin-dependent splitting of light beam. The polarization state generator (PSG) and the polarization state analyzer (PSA) units comprising of a fixed linear polarizer and a rotatable quarter waveplate, are used to generate and analyze desirable polarization states of light. $L_1$, $L_2$, $L_3$: Lenses; $P_1$, $P_2$: linear polarizers; $QWP_1$, $QWP_2$: quarter waveplates, SLM: spatial light modulator, $M_1$, $M_2$: mirrors, PH: pinhole, A: aperture here. While studying the spin specific beam shift, the PSG unit was used to sequentially generate RCP and LCP polarization states and the PSA unit was removed. For observing the spin-dependent splitting of light, the PSG was used to generate linear polarization state, and the transmitted beam was sequentially analyzed via the RCP and LCP analyzer states of the PSA. The same system was also employed to record Mueller matrices from the SLM.

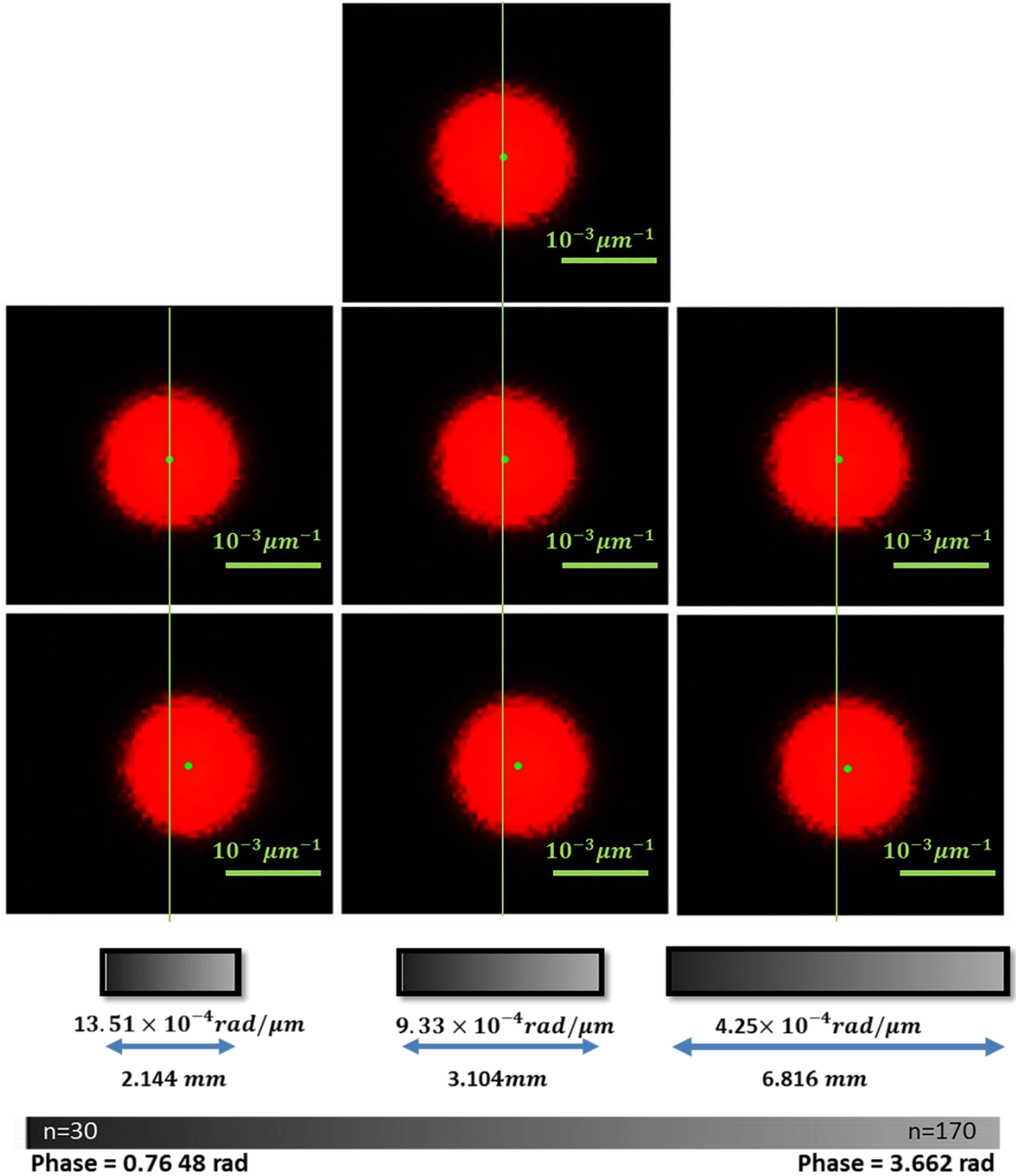

**Figure 2:** Experimental observation of tunable spin specific beam shift. Transverse momentum distribution ($\frac{k_{x/y}}{2\pi}$ μm$^{-1}$) of the transmitted beam for input LCP (middle panel) and RCP (bottom panel) states, for three representative spatial gradients of grey levels $\frac{dn}{d\xi=x/y}=$ 0.0205, 0.0451, 0.0653 bit/μm (right to left). Grey levels (values $n$= 30-170) are shown using color bar and the spatial dimensions over which these were applied, are noted. Result for a uniform grey level distribution is displayed as reference (top panel). While, the beam centroid for input RCP state exhibits large and systematic shift with increasing $\frac{dn}{d\xi=x/y}$, that for LCP state does not exhibit any appreciable shift, demonstrating spin specific beam shift and its tunability. The grey level dependence of total phase experienced by RCP state $\Phi_{tot}^{+}(n)$ (shown using color bar) and the noted phase gradients are based on determination of geometric and dynamical phases, results of which are presented subsequently.

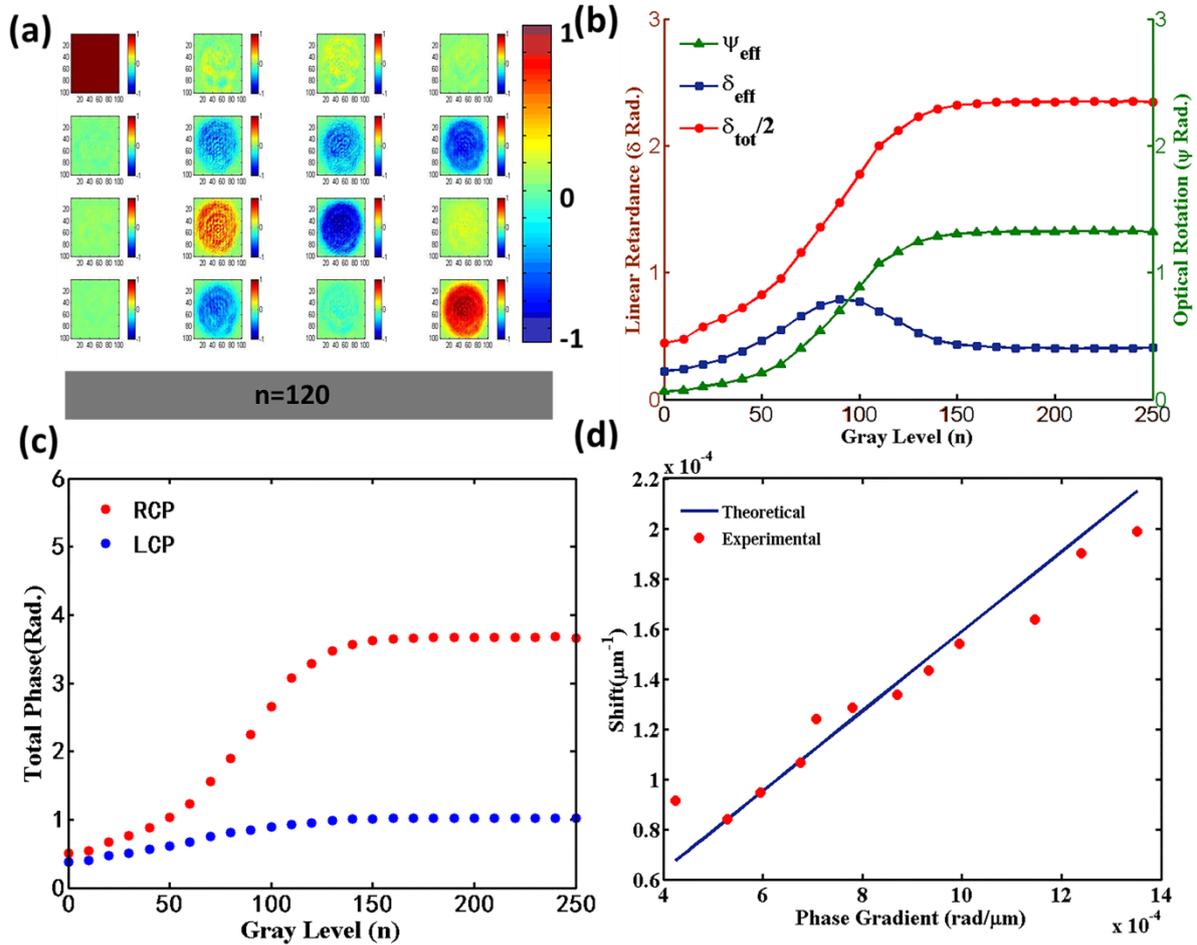

**Figure 3:** Experimental determination of PB geometric phase and dynamical phase of light. *(a)* Illustrative example of Mueller matrix recorded from the SLM having a uniform grey level addressing of *n* = 120. *(b)* The variation of the polarization parameters, effective linear retardance $\delta_{eff}(n)$ (blue square), optical rotation $\psi_{eff}(n)$ (green triangle) and total retardance $\delta_{tot}(n)$ (red circle) (line is guide for eye). *(c)* The corresponding variation of the total (dynamical + geometric) phase $[\Phi_{tot}^{\pm}(n) = \Phi_d(n) + \Phi_g^{\pm}(n)]$ for input RCP ($\Phi_{tot}^{+}(n)$, red circle) and LCP ($\Phi_{tot}^{-}(n)$, blue circle) states. *(d)* Comparison of the experimental shifts for RCP state (red circle) and corresponding theoretical predictions (blue line). Gradual increase of $\Phi_{tot}^{+}(n)$ (for *n* = 30 – 170 used in the experiments of Figure 2) for RCP state, corresponding negligible variation for LCP state and the agreement between the experiments and theory provide concrete evidence of the underlying principle of the spin specific beam shift.

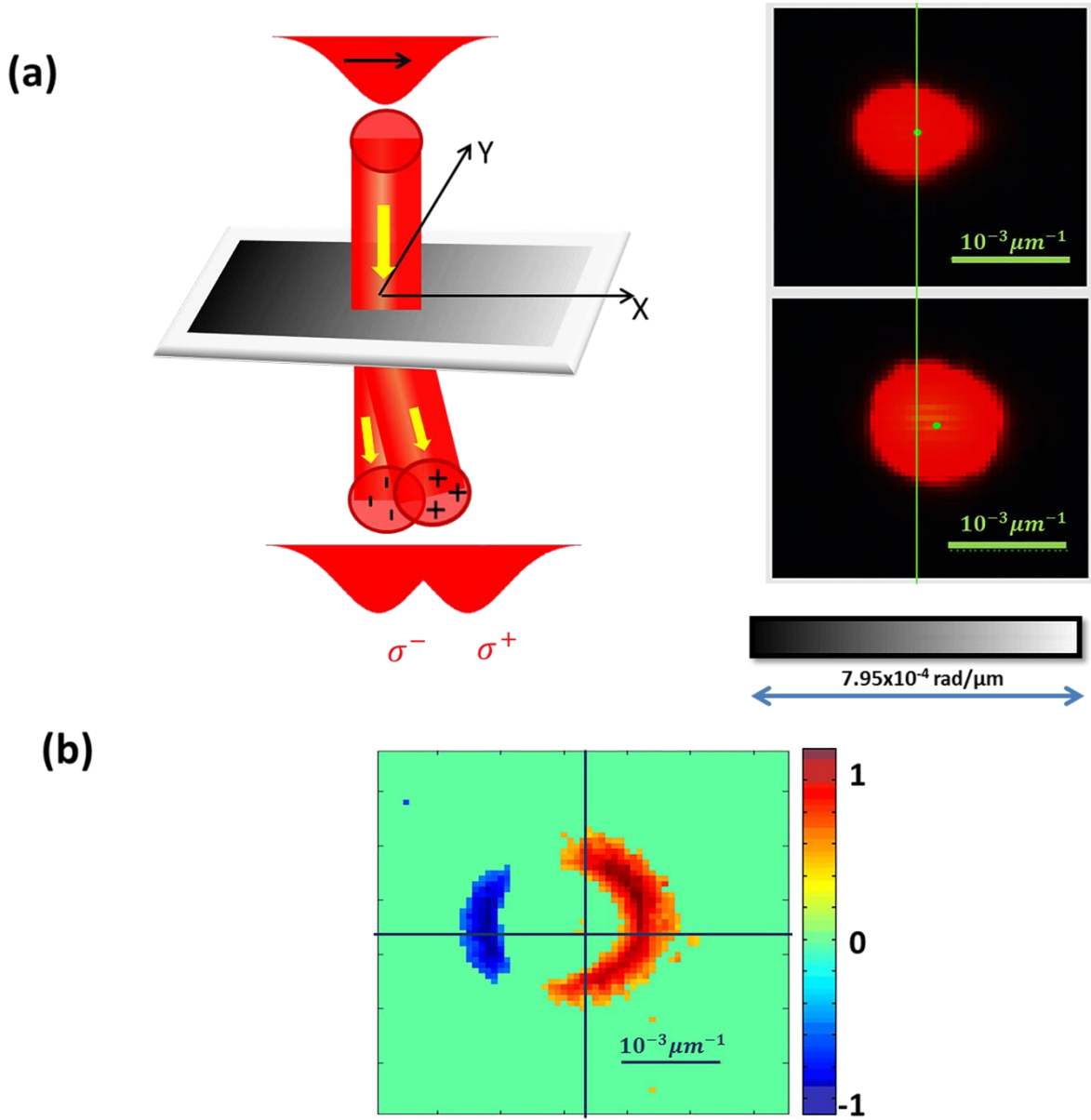

**Figure 4:** *(a)* Spin dependent splitting of input linearly polarized light in an inhomogeneous anisotropic medium exhibiting nearly equal spatial gradient of geometric and dynamical phases. One of the constituent circular polarization mode (RCP, noted as $\sigma^+$) experiences a large and tunable momentum domain shift (manifested as a shift of the beam centroid in the detection plane, shown in bottom panel), the other orthogonal mode (LCP, $\sigma^-$, top panel) evolves in the same trajectory. The results are displayed for a spatial gradient of the total phase of $\frac{d\Phi_{tot}^+(x)}{dx} = 7.95\times10^{-4}$ rad /μm (as experienced by the RCP mode). *(b)* The spin separation is shown by the spatial distribution of the circular polarization descriptor Stokes Vector element *V/I* (*V* is the difference in intensities between the RCP and LCP components and *I* is the sum of the two, the total intensity).